\documentstyle[12pt]{article}
\vspace{1.8cm} \oddsidemargin 0pt \evensidemargin 0pt  \topmargin
-1.5cm \footheight 26pt \footskip 10mm

\begin{document}
\baselineskip 20pt
\begin{center}
\baselineskip=24pt {\Large\bf  Schr\"{o}dinger equation of general
potential}

\vspace{1cm} \centerline{Xiang-Yao Wu$^{a}$
\footnote{E-mail:wuxy2066@163.com }, Xiao-Jing Liu$^{a}$, Yi-Heng
Wu$^{a}$, Qing-Cai Wang$^{a}$ and Yan Wang$^{a}$} \vskip 10pt
\noindent{\footnotesize a. \textit{Institute of Physics, Jilin
Normal University, Siping 136000, China}}

\end{center}
\date{}
\renewcommand{\thesection}{Sec. \Roman{section}} \topmargin 10pt
\renewcommand{\thesubsection}{ \arabic{subsection}} \topmargin 10pt
{\vskip 5mm
\begin {minipage}{140mm}
\centerline {\bf Abstract} \vskip 8pt
\par
\indent\\

\hspace{0.3in}It is well known that the Schr\"{o}dinger equation
is only suitable for the particle in common potential
$V(\vec{r},t)$. In this paper, a general Quantum Mechanics is
proposed, where the Lagrangian is the general form. The new
quantum wave equation can describe the particle which is in
general potential $V(\vec{r}, \dot{\vec{r}}, t)$.
We think these new quantum wave equations can be applied in many fields.


\end {minipage}

\newpage
\section * {1. Introduction }

\hspace{0.3in}It is well known that quantum mechanics (QM)
acquired its final formulation in 1925-1926 through fundamental
papers of Schr\"{o}dinger and Heisenberg. Originally these papers
appeared as two independent views of the structure of quantum
mechanics, but in 1927 Schr\"{o}dinger established their
equivalence, and since then one or the other of the papers
mentioned have been used to analyze quantum mechanical systems,
depending on which method gave the most convenient way of solving
the problem. Thus the existence of alternative procedures to solve
a given problem can be quite fruitful in deriving solutions of it.
Quantum Mechanics has become one of the most important foundations
of physics, and achieved great success, physicists had begun to
consider the possibility to generalize the traditional framework
of it[1]. Up to now, although various generalizations of QM are
all not very useful or successful, this kind of attempts have
never stopped[2-6].

In the 1940's Richard Feynman, and later many others, derived a
propagator for quantum mechanical problems through a path
integration procedure[7-9]. In contrast with the Hamiltonian
emphasis in the original formulation of quantum mechanics,
Feynmans approach could be referred to as Lagrangian and it
emphasized the propagator $K(x, t; x', t')$ which takes the wave
function $\psi(x', t')$ at the point $x'$ and time $t'$ to the
point $x$ at time $t$. While this propagator could be derived by
the standard methods of quantum mechanics. Feynman invented a
procedure by summing all time dependent paths connecting points
$x, x'$ and this became an alternative formulation of quantum
mechanics whose results coincided with the older version when all
of them where applicable, but also became relevant for problems
that the original methods could not solve.

At present, quantum mechanics is only suitable for common
potential $V(\vec{r}, t)$, but not suitable for general potential
$V(\vec{r}, \dot{\vec{r}}, t)$. In this paper, we extend the
Schr\"{o}dinger equation to suit general potential.

\section * {2. The Schr\"{o}dinger wave equation for general
Lagrangian}

\hspace{0.3in}It's well known that the Schr\"{o}dinger wave
equation can be obtained by Feynmal path integral. For the
Lagrangian function
\begin{equation}
L(\vec{r}, \dot{\vec{r}}, t)=\frac{1}{2} m
\dot{\overrightarrow{r}} ^{2}-V(r, t).
\end{equation}
For each path $x(t)$ connecting $(x_{i}, t_{i})$ and $(x_{f},
t_{f})$, calculate the action $S[\vec{r} (t)]$ defined by
\begin{equation}
S[\vec{r}(t)]=\int_{t_{i}}^{t_{f}}L(\vec{r}(t), \dot{\vec{r}}(t),
t)dt,
\end{equation}
with the help of the formula of Feynman path integral
\begin{equation}
\psi(\vec{r}_{2}, t_{2})=\int d\vec{r}_{1} K(\vec{r}_{2}t_{2};
\vec{r}_{1}t_{1}) \psi(\vec{r}_{1},t_{1}),
\end{equation}
where
\begin{equation}
K(\vec{r}_{2}t_{2}; \vec{r}_{1}t_{1})=\int
exp[iS[\vec{r}(t)]/\hbar]D[\vec{r}(t)],
\end{equation}
we can obtain the Schr\"{o}dinger wave equation
\begin{equation}
i\hbar\frac{\partial}{\partial t}\psi(\vec{r},
t)=[-\frac{\hbar}{2m}\nabla^{2}+V(r, t)]\psi(\vec{r}, t).
\end{equation}
In the following, we will give the general Schr\"{o}dinger wave
equation for the general Lagrangian function

\begin{equation}
L(x, \dot{x},
t)=a(t)\dot{x}^{2}+b(t)x\dot{x}+c(t)x^{2}+d(t)\dot{x}+f(t)x+g(t),
\end{equation}
we have already found that as a consequence of Eq. (3), we have
the equation
\begin{equation}
\psi(x_{2}, t_{2})=\int_{-\infty}^{\infty}K(x_{2}, t_{2}; x_{1},
t_{1})\psi(x_{1},t_{1})dx_{1},
\end{equation}
Eq. (7) gives the wave function at a time $t_{2}$ in terms of the
wave function at a time $t_{1}$. In order to obtain the
differential equation, we apply this relationship in the special
case that the time $t_{2}$ differs only by an infinitesimal
interval $\varepsilon$ from $t_{1}$. The propagator $K(x_{2},
t_{2}; x_{1}, t_{1})$ is proportional to the exponential of
$\frac{i}{\hbar}$ times the action for the interval $t_{1}$ to
$t_{2}$. For a short interval $\varepsilon$ the action is
approximately $\varepsilon$ times the Lagrangian for this
interval, we have
\begin{equation}
\psi(x, t+\varepsilon)=C\int_{-\infty}^{\infty}exp
[\varepsilon\frac{i}{\hbar}L(\frac{x-x'}{\varepsilon},
\frac{x+x'}{2})]\psi(x', t)dx',
\end{equation}
substituting Eq.(6) into (8), one can obtain
\begin{eqnarray}
\psi(x,t+\varepsilon)
&=&C\int^{\infty}_{-\infty}\exp[\frac{i\varepsilon}{h}
(a(t)(\frac{x-x'}{\varepsilon})^{2}+b(t)\frac{x+x'}{2}\frac{x-x'}{\varepsilon}
+c(t)(\frac{x+x'}{2})^{2}\nonumber\\
&&+d(t)\frac{x-x'}{\varepsilon}+ f(t)\frac{x+x'}{2}+g(t))]
\psi(x',t)dx'\nonumber\\
&=&C\int^{\infty}_{-\infty}\exp[\frac{i}{h}
(a(t)\frac{(x-x')^{2}}{\varepsilon}+b(t)\frac{x+x'}{2} (x-x')
+c(t)\varepsilon(\frac{x+x'}{2})^{2}\nonumber\\
&&+d(t)(x-x')+ f(t)\varepsilon\frac{x+x'}{2}+g(t)\varepsilon)]
\psi(x',t)dx'.
\end{eqnarray}
The quantity $\frac{(x-x')^{2}}{\varepsilon}$ appear in the
exponent of the first factor. It is clear that if $x'$ is
appreciably different from $x$, this quantity is very large and
the exponential consequently oscillates very rapidly as $x'$
varies, when this factor oscillates rapidly, the integral over
$x'$ gives a very small value. Only if $x'$ is near $x$ do we get
important contributions. For this reason we make the substitution
$x'=x+\eta$ with the expectation that appreciable contribution to
the integral will occur only for small $\eta$, we obtain
\begin{eqnarray}
\psi(x,t+\varepsilon)&=&C\int^{\infty}_{-\infty}\exp[\frac{i\eta^{2}}{\hbar\varepsilon}a(t)]\cdot
\exp[\frac{i}{\hbar}(x+\frac{\eta}{2})(-\eta)b(t)]
\cdot\exp[\frac{i\varepsilon}{\hbar}(x+\frac{\eta}{2})^{2}c(t)]\nonumber\\
&&\cdot\exp[\frac{i}{\hbar}(-\eta)d(t)]
\cdot\exp[\frac{i\varepsilon}{\hbar}(x+\frac{\eta}{2})f(t)]
\cdot\exp[\frac{i\varepsilon}{\hbar}g(t)]
\psi(x+\eta,t)d\eta\nonumber\\
&=&C\int^{\infty}_{-\infty}\exp[\frac{i{\eta}^{2}}{\hbar\varepsilon}a(t)]
\cdot\exp[-\frac{i\eta}{\hbar}b(t)x]
\cdot\exp[-\frac{i\eta^{2}}{2\hbar}b(t)]
\cdot\exp[\frac{i\varepsilon}{\hbar}c(t)x^{2}]
\nonumber\\
&& \cdot\exp[\frac{i\varepsilon}{\hbar}x\eta
c(t)]\cdot\exp[\frac{i\varepsilon\eta^{2}}{4\hbar}
c(t)]\cdot\exp[-\frac{i\eta}{\hbar}
d(t)]\cdot\exp[\frac{i\varepsilon}{\hbar}f(t)x]\nonumber\\
&& \cdot\exp[\frac{i\varepsilon\eta}{2\hbar}f(t)]
\cdot\exp[\frac{i\varepsilon}{\hbar}g(t)] \psi(x+\eta,t)d\eta.
\end{eqnarray}
The phase of the first exponential charges by the order of 1
radian when $\eta$ is of the order
$\sqrt{\frac{\hbar\varepsilon}{a(t)}}$, so that most of the
integral is contributed by values of $\eta$ in this order.
  We may expand $\psi$ in a power series, we need only keep terms
of order $\varepsilon$. This implies keeping second-order terms in
$\eta$. Expanding the left-hand side to first order in
$\varepsilon$ and the right-hand side to first order in
$\varepsilon$ and second order in $\eta$, we have

\begin{equation}
e^{-\frac{i\eta}{\hbar}b(t)x}=1-\frac{i\eta}{\hbar}b(t)x+\frac{1}{2}\frac{\eta^{2}}{\hbar^{2}}
{b^{2}(t)}x^{2},
\end{equation}
\begin{equation}
e^{-\frac{i\eta^{2}}{2\hbar}b(t)}=1-\frac{i\eta^{2}}{2\hbar}b(t),
\end{equation}
\begin{equation}
e^{\frac{i\varepsilon}{\hbar}c(t)x^{2}}=1+\frac{i\varepsilon}{\hbar}c(t)x^{2},
\end{equation}
\begin{equation}
e^{\frac{i\varepsilon\eta^{2}}{4\hbar}c(t)}=1+\frac{i\varepsilon\eta^{2}}{4\hbar}c(t),
\end{equation}
\begin{equation}
e^{\frac{i\varepsilon\eta}{\hbar}c(t)x}=1+\frac{i\varepsilon\eta}{\hbar}c(t)x,
\end{equation}
\begin{equation}
e^{-\frac{i\eta}{\hbar}d(t)}=1-\frac{i\eta}{\hbar}d(t)+\frac{1}{2}\frac{\eta^{2}}{\hbar^{2}}
{d^{2}(t)},
\end{equation}
\begin{equation}
e^{\frac{i\varepsilon}{\hbar}f(t)x}=1+\frac{i\varepsilon}{\hbar}f(t)x,
\end{equation}
\begin{equation}
e^{\frac{i\varepsilon\eta}{2\hbar}f(t)}=1+\frac{i\varepsilon\eta}{2\hbar}f(t),
\end{equation}
\begin{equation}
e^{\frac{i\varepsilon}{\hbar}g(t)}=1+\frac{i\varepsilon}{\hbar}g(t),
\end{equation}
\begin{equation}
\psi(x+\eta,t)=\psi(x,t)+\eta\frac{{\partial}\psi}{\partial
x}+\frac{1}{2}\eta^{2}\frac{{\partial}^{2}\psi}{\partial x^{2}} ,
\end{equation}
and
\begin{eqnarray}
\psi(x,t)+\varepsilon\frac{{\partial}\psi}{\partial
t}=C\int^{\infty}_{-\infty}\exp[\frac{i\eta^{2}}{\hbar\varepsilon}a(t)]
(1-\frac{i\eta}{\hbar}b(t)x+\frac{1}{2}\frac{\eta^{2}}{\hbar^{2}}b^{2}(t)x^{2})
(1-\frac{i\eta^{2}}{2\hbar}b(t))\nonumber\\
(1+\frac{i\varepsilon}{\hbar}c(t)x^{2})
(1+\frac{i\varepsilon\eta^{2}}{4\hbar}c(t))
(1+\frac{i\varepsilon\eta x}{\hbar}c(t))
(1-\frac{i\eta}{\hbar}d(t)+\frac{1}{2}\frac{\eta^{2}}{\hbar^{2}}d^{2}(t))\nonumber\\
(1+\frac{i\varepsilon}{\hbar}f(t)x)
(1+\frac{i\varepsilon\eta}{2\hbar}f(t))
(1+\frac{i\varepsilon}{\hbar}g(t))
(\psi(x,t)+\eta\frac{\partial\psi}{\partial
{x}}+\frac{1}{2}\eta^{2}\frac{{\partial^{2}}\psi}{\partial
x^{2}})d\eta,
\end{eqnarray}
when $\varepsilon\rightarrow0$ and $\eta\rightarrow0$, the Eq.
(21) becomes
\begin{equation}
\psi(x,t)=C\int^{\infty}_{-\infty}\exp[\frac{i\eta^{2}}{\hbar\varepsilon}a(t)]\psi(x,t)d\eta,
\end{equation}
and the constant $C$ is
\begin{eqnarray}
C=\frac{1}{\int^{\infty}_{-\infty}\exp[\frac{i\eta^{2}}{\hbar\varepsilon}a(t)]d\eta}
=\sqrt{\frac{a(t)}{i\pi\hbar\varepsilon}} .
\end{eqnarray}
In order to evaluate the right-hand side of Eq. (21), we shall
have to use three integrals
\begin{equation}
\int_{-\infty}^{\infty}e^{\frac{i\eta^{2}}{\hbar\varepsilon}a(t)}\eta
d\eta=0,
\end{equation}
\begin{equation}
\int^{\infty}_{-\infty}e^{\frac{i\eta^{2}}{\hbar\varepsilon}a(t)}\eta^{2}
d\eta=\frac{ i\hbar\varepsilon}{2a(t)}\sqrt{\frac{i\pi
\hbar\varepsilon}{a(t)}},
\end{equation}
\begin{equation}
\int^{\infty}_{-\infty}e^{\frac{i\eta^{2}}{\hbar\varepsilon}a(t)}\eta^{4}
d\eta=\frac{3}{4}(i\hbar\varepsilon)^{2}\frac{\sqrt{i\pi
\hbar\varepsilon}}{a(t)^{\frac{5}{2}}}.
\end{equation}
In Eq. (21), we can easily find the terms
$\frac{i\varepsilon\eta^{2}}{4\hbar}c(t)$, $\frac{i\varepsilon\eta
x}{\hbar}c(t)$ and $\frac{i\varepsilon\eta}{2\hbar}f(t)$ integral
are zero or $O(\varepsilon^{2})$ from Eqs. (24)-(26), and they can
be neglected in Eq. (21). The Eq. (21) becomes
\begin{eqnarray}
\psi(x,t)+\varepsilon\frac{{\partial}\psi}{\partial
t}&=&C\int^{\infty}_{-\infty}\exp[\frac{i\eta^{2}}{\hbar\varepsilon}a(t)]
(1-\frac{i\eta}{\hbar}b(t)x+\frac{1}{2}\frac{\eta^{2}}{\hbar^{2}}b^{2}(t)x^{2})
(1-\frac{i\eta^{2}}{2\hbar}b(t))\nonumber\\
&&(1+\frac{i\varepsilon}{\hbar}c(t)x^{2})
(1-\frac{i\eta}{\hbar}d(t)+\frac{1}{2}\frac{\eta^{2}}{\hbar^{2}}d^{2}(t))
(1+\frac{i\varepsilon}{\hbar}f(t)x)\nonumber\\&&
(1+\frac{i\varepsilon}{\hbar}g(t))
(\psi(x,t)+\eta\frac{\partial\psi}{\partial
{x}}+\frac{1}{2}\eta^{2}\frac{{\partial^{2}}\psi}{\partial
x^{2}})d\eta,
\end{eqnarray}

we defined
\begin{eqnarray}
I&=&(1-\frac{i\eta}{\hbar}b(t)x+\frac{1}{2}\frac{\eta^{2}}{\hbar^{2}}b^{2}(t)x^{2})
(1-\frac{i\eta^{2}}{2\hbar}b(t))(1+\frac{i\varepsilon}{\hbar}c(t)x^{2})\nonumber\\
&&(1-\frac{i\eta}{\hbar}d(t)+\frac{1}{2}\frac{\eta^{2}}{\hbar^{2}}d^{2}(t))
(1+\frac{i\varepsilon}{\hbar}f(t)x)(1+\frac{i\varepsilon}{\hbar}g(t))\nonumber\\
&=&(1-\frac{i\eta}{\hbar}b(t)x+\frac{1}{2}\frac{\eta^{2}}{\hbar^{2}}b^{2}(t)x^{2}
-\frac{i\eta^{2}}{2\hbar}b(t)-\frac{\eta^{3}}{2\hbar^{2}}b^{2}(t)x
-\frac{i\eta^{4}}{4\hbar^{3}}b^{3}(t)x^{2})(1+\frac{i\varepsilon}{\hbar}c(t)x^{2})\nonumber\\
&&(1-\frac{i\eta}{\hbar}d(t)+\frac{1}{2}\frac{\eta^{2}}{\hbar^{2}}d^{2}(t))
(1+\frac{i\varepsilon}{\hbar}g(t)+\frac{i\varepsilon}{\hbar}f(t)x
-\frac{\varepsilon^{2}}{\hbar^{2}}f(t)g(t)x).
\end{eqnarray}
In Eq. (28), the terms $\frac{i\eta^{3}}{2\hbar^{2}}b^{2}(t)x$,
$\frac{i\eta^{4}}{4\hbar^{3}}b^{2}(t)x^{2}$ and
$\frac{x\varepsilon^{2}}{\hbar^{2}}f(t)g(t)$ integral are either
zero or $O(\varepsilon^{2})$ from Eqs. (24)-(26), and they can be
neglected in Eq. (27). The function $I$ becomes
\begin{eqnarray}
I&=&(1-\frac{i\eta}{\hbar}b(t)x+\frac{1}{2}\frac{\eta^{2}}{\hbar^{2}}b^{2}(t)x^{2}
-\frac{i\eta^{2}}{2\hbar}b(t))(1+\frac{i\varepsilon}{\hbar}c(t)x^{2})\nonumber\\
&&(1-\frac{i\eta}{\hbar}d(t)+\frac{1}{2}\frac{\eta^{2}}{\hbar^{2}}d^{2}(t))
(1+\frac{i\varepsilon}{\hbar}g(t)+\frac{i\varepsilon}{\hbar}f(t)x)\nonumber\\
&=&(1-\frac{i\eta}{\hbar}b(t)x+\frac{1}{2}\frac{\eta^{2}}{\hbar^{2}}b^{2}(t)x^{2}
-\frac{i\eta^{2}}{2\hbar}b(t))
(1-\frac{i\eta}{\hbar}d(t)+\frac{1}{2}\frac{\eta^{2}}{\hbar^{2}}d^{2}(t))\nonumber\\
&&(1+\frac{i\varepsilon}{\hbar}g(t)
+\frac{i\varepsilon}{\hbar}f(t)x+\frac{i\varepsilon}{\hbar}c(t)x^{2}
-\frac{\varepsilon^{2}}{\hbar^{2}}c(t)g(t)x^{2}
-\frac{\varepsilon^{2}}{\hbar^{2}}c(t)f(t)x^{3}).
\end{eqnarray}
In Eq. (29), we neglect terms including $\varepsilon^{2}, \eta^{3}
$ and $\eta^{4}$, and $I$ can be written as
\begin{eqnarray}
I&=&(1-\frac{i\eta}{\hbar}b(t)x+\frac{1}{2}\frac{\eta^{2}}{\hbar^{2}}b^{2}(t)x^{2}
-\frac{i\eta^{2}}{2\hbar}b(t)-\frac{i\eta}{\hbar}d(t)
-\frac{\eta^{2}}{\hbar^{2}}b(t)d(t)x+\frac{1}{2}\frac{\eta^{2}}{\hbar^{2}}d^{2}(t))\nonumber\\
&&(1+\frac{i\varepsilon}{\hbar}g(t)+\frac{i\varepsilon}{\hbar}f(t)x
+\frac{i\varepsilon}{\hbar}c(t)x^{2})\nonumber\\
&=&(1-\frac{i\eta}{\hbar}(b(t)x+d(t))
+\frac{1}{2}\frac{\eta^{2}}{\hbar^{2}}(b^{2}(t)x^{2}-i\hbar
b(t)+d^{2}(t)-2b(t)d(t)x))\nonumber\\
&&(1+\frac{i\varepsilon}{\hbar}(g(t)+f(t)x
+c(t)x^{2}))\nonumber\\
&=&(1-\frac{i\eta}{\hbar}A(x,t)+\frac{1}{2}\frac{\eta^{2}}{\hbar^{2}}B(x,t))
(1+\frac{i\varepsilon}{\hbar}C(x,t)),
\end{eqnarray}
where the functions $A(x, t)$, $B(x, t)$ and $C(x, t)$ are
\begin{eqnarray}
A(x,t)&=& b(t)x+d(t), \nonumber\\
B(x,t)&=&b^{2}(t)x^{2}-i\hbar
b(t)+d^{2}(t)-2b(t)d(t)x,\nonumber\\
C(x,t)&=&g(t)+f(t)x+c(t)x^{2},
\end{eqnarray}
substituting Eq. (30) into (27), we have
\begin{eqnarray}
\psi(x,t)+\varepsilon\frac{{\partial}\psi}{\partial t}&=&
C\int^{\infty}_{-\infty}e^{\frac{i\eta^{2}}{\hbar\varepsilon}a(t)}
(1-\frac{i\eta}{\hbar}A(x,t)+\frac{1}{2}\frac{\eta^{2}}{\hbar^{2}}B(x,t))(1
+\frac{i\varepsilon}{\hbar}C(x, t)) \nonumber\\&&
\cdot(\psi(x,t)+\eta\frac{{\partial}\psi}{\partial
x}+\frac{1}{2}\eta^{2}\frac{{\partial}^{2}\psi}{\partial
x^{2}})d\eta\nonumber\\ &=&
C\int^{\infty}_{-\infty}e^{\frac{i\eta^{2}}{\hbar\varepsilon}a(t)}
(1-\frac{i\eta}{\hbar}A(x,t)+\frac{1}{2}\frac{\eta^{2}}{\hbar^{2}}B(x,t)
+\frac{i\varepsilon}{\hbar}C(x,t)
+\frac{\eta\varepsilon}{\hbar^{2}}A(x,t)\nonumber\\&& \cdot C(x,t)
+\frac{i}{2}\frac{\eta^{2}\varepsilon}{\hbar^{3}}B(x,t)C(x,t))\cdot(\psi(x,t)+\eta\frac{{\partial}\psi}{\partial
x}+\frac{1}{2}\eta^{2}\frac{{\partial}^{2}\psi}{\partial
x^{2}})d\eta.
\end{eqnarray}
In Eq. (32), the terms including $\eta \varepsilon$ and
$\eta^{2}\varepsilon$ integral are either zero or
$O(\varepsilon^{2})$, and they can be neglected in Eq. (32). The
Eq. (32) becomes
\begin{eqnarray}
\psi(x,t)+\varepsilon\frac{{\partial}\psi}{\partial t}&=&
C\int^{\infty}_{-\infty}e^{\frac{i\eta^{2}}{\hbar\varepsilon}a(t)}
(1-\frac{i\eta}{\hbar}A(x,t)+\frac{1}{2}\frac{\eta^{2}}{\hbar^{2}}B(x,t)
+\frac{i\varepsilon}{\hbar}C(x,t))\nonumber\\
&&\cdot(\psi(x,t)+\eta\frac{{\partial}\psi}{\partial
x}+\frac{1}{2}\eta^{2}\frac{{\partial}^{2}\psi}{\partial
x^{2}})d\eta.
\end{eqnarray}
In Eq. (33), the terms including $\eta$, $\eta^{3}$ and $\eta^{4}$
integral are either zero or $O(\varepsilon^{2})$, and they can be
neglected in Eq. (33). The Eq. (33) becomes

\begin{eqnarray}
\psi(x,t)+\varepsilon\frac{{\partial}\psi}{\partial
t}&=&C\int^{\infty}_{-\infty}e^{\frac{i\eta^{2}}{\hbar\varepsilon}a(t)}
(\psi(x,t) +\frac{1}{2}\frac{\eta^{2}}{\hbar^{2}}B(x,t)\psi(x,t)
+\frac{i\varepsilon}{\hbar}C(x,t)\psi(x,t)\nonumber\\
&&-\frac{i\eta^{2}}{\hbar}A(x,t) \frac{{\partial}\psi}{\partial x}
+\frac{1}{2}\eta^{2}\frac{{\partial^{2}}\psi}{\partial
x^{2}})d\eta\nonumber\\
&=&\psi(x,t)+\frac{1}{2\hbar^{2}}B(x,t)\psi(x,t)\cdot
C\int^{\infty}_{-\infty}e^{\frac{i\eta^{2}}{\hbar\varepsilon}a(t)}\eta^{2}d\eta
+\frac{i\varepsilon}{\hbar}C(x,t)\psi(x,t) \nonumber\\&&\cdot
C\int^{\infty}_{-\infty}e^{\frac{i\eta^{2}}{\hbar\varepsilon}a(t)}d\eta
-\frac{i}{\hbar}A(x,t)\frac{{\partial}\psi}{\partial x}\cdot
C\int^{\infty}_{-\infty}e^{\frac{i\eta^{2}}{\hbar\varepsilon}a(t)}\eta^{2}d\eta
+\frac{1}{2}\frac{{\partial}^{2}\psi}{\partial x^{2}}
\nonumber\\&& \cdot
C\int^{\infty}_{-\infty}e^{\frac{i\eta^{2}}{\hbar\varepsilon}a(t)}\eta^{2}d\eta
\nonumber\\
&=&\psi(x,t)+\frac{1}{2\hbar^{2}}B(x,t)\psi(x,t)
\frac{i\hbar\varepsilon}{2a(t)}+\frac{i\varepsilon}{\hbar}C(x,t)\psi(x,t)
\nonumber\\&& -\frac{i}{\hbar}A(x,t)\frac{{\partial}\psi}{\partial
x}\frac{i\hbar\varepsilon}{2a(t)}
+\frac{1}{2}\frac{{\partial}^{2}\psi}{\partial
x^{2}}\frac{i\hbar\varepsilon}{2a(t)}.
\end{eqnarray}
Equating the coefficient of powers of $\varepsilon$, we have
\begin{equation}
\frac{{\partial}\psi(x,t)}{\partial t}=\frac{i}{4\hbar
a(t)}B(x,t)\psi(x,t)+\frac{i}{\hbar}C(x,t)\psi(x,t)
+\frac{1}{2a(t)}A(x,t)\frac{{\partial}\psi}{\partial x}
+\frac{i\hbar}{4a(t)}\frac{{\partial}^{2}\psi}{\partial x^{2}},
\end{equation}
multiplied the coefficient of $i\hbar$, we have
\begin{equation}
i\hbar\frac{{\partial}\psi(x,t)}{\partial t}=-\frac{1}{4
a(t)}B(x,t)\psi(x,t)-C(x,t)\psi(x,t)+i\frac{\hbar}{2 a(t)}
A(x,t)\frac{{\partial}\psi}{\partial x}-
\frac{\hbar^{2}}{4a(t)}\frac{{\partial}^{2}\psi}{\partial x^{2}},
\end{equation}
substituting the function $A(x,t)$, $B(x,t)$ and $C(x,t)$ into Eq.
(36), we obtain the general Schr\"{o}dinger equation
\begin{eqnarray}
&&i\hbar\frac{{\partial}\psi(x,t)}{\partial
t}=-\frac{\hbar^{2}}{4a(t)}\frac{{\partial^{2}}\psi}{\partial
x^{2}}+i\frac{\hbar}{2
a(t)}(b(t)x+d(t))\frac{{\partial}\psi}{\partial x}\nonumber\\&&
-(\frac{b^{2}(t)}{4a(t)}x^{2}-\frac{i\hbar
b(t)}{4a(t)}+\frac{d^{2}(t)}{4a(t)}-\frac{
b(t)d(t)}{2a(t)}x+g(t)+f(t)x+c(t)x^{2})\psi(x,t).
\end{eqnarray}
The Eq. (37) is a general Schr\"{o}dinger equation for
one-dimensional. We
can discuss Eq. (37) as follows:\\
(a) When $b(t)=c(t)=d(t)=g(t)=0$, the Lagrangian function (6) is
\begin{equation}
L(x, \dot{x}, t)=a(t)\dot{x}^{2}+f(t)x,
\end{equation}
Eq. (37) becomes
\begin{eqnarray}
i\hbar\frac{{\partial}\psi(x,t)}{\partial
t}&=&-\frac{\hbar^{2}}{4a(t)}\frac{{\partial^{2}}\psi(x,
t)}{\partial x^{2}}-f(t)x\psi(x,t).
\end{eqnarray}
(b) When $b(t)=d(t)=f(t)=g(t)=0$, the Lagrangian function (6) is
\begin{equation}
L(x, \dot{x}, t)=a(t)\dot{x}^{2}+c(t)x^{2},
\end{equation}
Eq. (37) becomes
\begin{eqnarray}
i\hbar\frac{{\partial}\psi(x,t)}{\partial
t}&=&-\frac{\hbar^{2}}{4a(t)}\frac{{\partial^{2}}\psi(x,
t)}{\partial x^{2}}-c(t)x^{2}\psi(x,t).
\end{eqnarray}
(c) When $c(t)=d(t)=f(t)=g(t)=0$, the Lagrangian function (6) is
\begin{equation}
L(x, \dot{x}, t)=a(t)\dot{x}^{2}+b(t)x\cdot\dot{x},
\end{equation}
Eq. (37) becomes
\begin{eqnarray}
i\hbar\frac{{\partial}\psi(x,t)}{\partial
t}&=&-\frac{\hbar^{2}}{4a(t)}\frac{{\partial^{2}}\psi}{\partial
x^{2}}+i\hbar\frac{b(t)x}{2 a(t)}\frac{{\partial}\psi}{\partial x}
-(\frac{b^{2}(t)}{4a(t)}x^{2}-\frac{i\hbar b(t)}{4a(t)})\psi(x,t).
\end{eqnarray}
(d) When $b(t)=c(t)=f(t)=g(t)=0$, the Lagrangian function (6) is
\begin{equation}
L(x, \dot{x}, t)=a(t)\dot{x}^{2}+d(t)\dot{x},
\end{equation}
Eq. (37) becomes
\begin{eqnarray}
i\hbar\frac{{\partial}\psi(x,t)}{\partial
t}&=&-\frac{\hbar^{2}}{4a(t)}\frac{{\partial^{2}}\psi(x,
t)}{\partial x^{2}}+i\hbar\frac{d(t)}{2
a(t)}\frac{{\partial}\psi(x, t)}{\partial x}
-\frac{d^{2}(t)}{4a(t)}\psi(x,t).
\end{eqnarray}
For the general Lagrangian function in three-dimensional, it is
\begin{equation}
L(\vec{r}, \dot{\vec{r}},
t)=a(t){\dot{\overrightarrow{r}}}^{2}+b(t)\vec{r}\cdot\dot{\vec{r}}+c(t)\overrightarrow{r}^{2}+g(t),
\end{equation}
We can easily obtain the general Schr\"{o}dinger equation for
three-dimensional, it is
\begin{eqnarray}
i\hbar\frac{{\partial}\psi(\vec{r},t)}{\partial
t}&=&-\frac{\hbar^{2}}{4a(t)}\nabla^{2}\psi(\vec{r},
t)+i\frac{\hbar}{2 a(t)}b(t)\vec{r}\cdot\nabla\psi(\vec{r},
t)\nonumber\\&&
-(\frac{b^{2}(t)}{4a(t)}\overrightarrow{r}^{2}-\frac{i\hbar
b(t)}{4a(t)}+c(t)\overrightarrow{r}^{2}+g(t))\psi(\vec{r},t).
\end{eqnarray}
Eq. (47) is a general Schr\"{o}dinger equation, which is suitable
for a general potential. From the equation, we can study the
general system.

\section * {3. Conclusion}

\hspace{0.3in}We know Schr\"{o}dinger equation is quantum wave
equation, which is only suitable for a particle in common
potential $V(\vec{r}, t)$. When a particle is in general potential
$V(\vec{r}, \dot{\vec{r}}, t)$, we need new quantum wave equation.
In this paper, we apply the approach of path integral to obtain
the general Schr\"{o}dinger equation, which is suitable for the
general potential. We think the general Schr\"{o}dinger equation
will be used widely in many fields.


\newpage
\begin{thebibliography}{99}
\bibitem{s1}
P. Jordan, Z. Phys. {\bf 80} (1933) 285; P. Jordan and E. P.
Wigner, Ann. Math. {\bf 35} (1934) 29.
\bibitem{s2}
D. Finkelstein et al., Notes on Quaternion Quantum Mechanics
(CERN, Report 59-7), in C. Hoker, ed. Logico-Algebraic-Approach to
Quantum Mechanics $\coprod$ (Reidel, Dordrecht, 1979).
\bibitem{s3}
D. Finkelstein et al., J. Math. Phys. {\bf 3} (1962) 207; ibid.
{\bf 4} (1963) 788
\bibitem{s4}
S. De Leo and K. Abdel-Khalek, Prog. Theor. Phys. {\bf 96} (1996)
823.
\bibitem{s5}
D. Minic and C. -H. Tze, Phys. Lett. B {\bf 581} (2004) 111.
\bibitem{s6}
Fred Riewe, Phys. Rev. E {\bf 55} (1997) 3581.
\bibitem{s7}
R. P. Feynman, Rev. Mod. Phys., {\bf 20}, (1948) 367.
\bibitem{s8}
R. P. Feynman and A. R. Hibbs, Quantum mechanics and path
integrals, McGraw-Hill Book Co.New York 1965
\bibitem{s9}
R. P. Feynman, Phys. Rev, {\bf 80}, 440 (1950)
\end{thebibliography}
\end{document}